# Scytale – An Evolutionary Cryptosystem

[1] Unnikrishnan Menon; [2] Atharva Hudlikar; [3] Divyani Panda

[1] Department of Electrical and Electronics Engineering, Vellore Institute of Technology
Vellore, Tamil Nadu 632014, India

[2] Department of Electrical and Electronics Engineering, Vellore Institute of Technology
Vellore, Tamil Nadu 632014, India

[3] Department of Electrical and Electronics Engineering, Vellore Institute of Technology
Vellore, Tamil Nadu 632014, India

**Abstract -** With the advent of quantum computing, and other advancements in computation and processing capabilities of modern systems, there arises a need to develop new trapdoor functions that will serve as the foundation for a new generation of encryption schemes. This paper explores the possibility of one such potential trapdoor function using concepts stemming from Reversible Cellular Automata (RCA) - specifically, the Critter's Rule set up in a Margolus Neighborhood. The proposed block encryption algorithm discusses how sensitive data can be manipulated and converted efficiently into a two dimensional sequence of bits, that can be iteratively evolved using the rules of the RCA and a private key to achieve a desirable level of encryption within a reasonable runtime. The performance benchmark and analysis results exemplify how well the proposed encryption algorithm stands against different forms of attacks.

## 1. Introduction

The establishment of Cryptology was well during the ancient times when both the Greeks and Romans practiced different forms of cryptography. [1] Being an old technique that is still being developed, cryptography nowadays, is being used all over the world daily by billions of people to secure data and information. Medieval cryptography was associated with the design and utilization of codes (also known as ciphers). In modern phraseology, these ciphers are called encryption schemes. Classical encryption schemes relied on symmetric key settings where a private key was shared between the communicators which was kept unknown to the eavesdroppers. The transmitter encrypts the plaintext using the shared key to obtain a ciphertext which is then decrypted by the receiver with the help of the same key. [2]

Various algorithms which have been developed so far to enumerate symmetric key cryptography are AES, DES, 3DES and Blowfish. [3]

IBM in 1997 developed a symmetric key cryptographic algorithm called Data Encryption Standard (DES) which operates on a block size of 64 bits. [4] In order to see how long it would take to decrypt a message, a series of challenges were sponsored, in which two organizations namely distributed.net and the Electronic Frontier Foundation (EFF) were successful in breaking DES. It took 84 days to break the encrypted text using a brute force attack in the DES I contest (1997). In 1998, the second challenge took around 3 days and the final DES III challenge in 1999 took only 22 hours and 15 minutes. [5]

Due to the inadequacy of DES, Triple Data Encryption(3DES) came into the picture which is an enhanced form of the former algorithm and has a key length of 192 bits. [4] In this process, DES encryption procedure is repeated three times. Nevertheless, 3DES still lacked the potential to protect data for a longer period against brute force attacks. [5].

In 1997, National Institute of Standards and Technology (NIST) organized an open competition for the replacement of DES algorithm which had to be efficient in not only hardware (like DES) but also software implementations. Finally, in October 2000, Rijndael was formally accepted as the proposed Advanced Encryption Standard (AES). The major strength of AES relies on the fact that it has options for different key lengths such as 128-bit, 192-bit or 256-bit key. This feature makes AES potentially stronger than that of the 56-bit key of DES. [5]

Currently, the world depends on AES which is virtually impenetrable by means of brute-force. However, no encryption system is wholly secure and researchers prodding into AES have found potential ways of cryptanalysis, one of them being a possible key-attack where the operation of a cipher in response to different keys is tracked. [6]

Cellular automata have intensive application in the currently emerging cryptography techniques used in both public and private key encryption systems. [9].

The advent of the first system of cellular automata happened in the late 1950s when Ulam and von Neumann created a method for calculating liquid motion which had the driving concept of taking into consideration a liquid as a group of discrete units, and calculate the motion of each based on its neighbors' behaviors.[10] It was not until Conway's Game of Life in 1970 that cellular automata gained popularity beyond academia.

This paper focuses on the application of two-dimensional reversible cellular automata suitable for symmetric key encryption systems. With the advent of high-end cyberattacks in the current scenario, it is of utmost importance to construct a more reliable and secure trapdoor function. The proposed RCA is a concept which lays the foundation of one such trapdoor function for further encryption systems.

## 2. Cellular Automata

A cellular automaton is a combination of cells with different states on a grid of specific shape and dimension that evolves through several discrete time steps based on a set of rules depending on the states of neighboring cells. These set of rules are then iteratively applied for the desired number of time steps. The concept was originally discovered in the 1940s by Stanislaw Ulam and Jon von Neumann. One of the most foundational properties of a cellular automaton is the type of grid on which it is computed. The simplest such "grid" is a one-dimensional line. In two dimensions, square, triangular, and hexagonal grids may be considered. Cellular Automata are initialized with one state with all $0$ s and a single $1$ at different locations. It can generate some fixed unique patterns.

A cellular automaton is a model of a system of "cell" objects. The three features or characteristics of cellular automata are:

a. The cells live on a grid. Typically, in one and two dimensions, though a cellular automaton can exist in any finite number of dimensions.
b. Each cell has a state. The number of state possibilities is typically finite. The simplest example has the two possibilities of 1 and 0 (otherwise referred to as "on" and "off" or "alive" and "dead").
c. Each cell has a neighborhood. This can be defined in any number of ways, but it is typically a list of adjacent cells.

## 3. Critter's Rule in Margolus Neighborhood

The Critter's Rule is a two-dimensional partitioning cellular automaton that operates in the Margolus Neighborhood [8], in which the grid or lattice is subdivided into blocks which are then independently updated based on certain rules.

In contrast to the local interaction of a CA which is in the form of $f : Q^n \rightarrow Q$, i.e., $n$ neighbors to 1 cell, a partitioning CA has a local interaction of the form, $f : Q^n \rightarrow Q^n$, i.e., $n$ neighbours to $n$ cells where each lattice is subdivided into a block consisting of $n$ cells. The partitioning scheme was introduced by Norman Margolus to study certain physical properties such as invertibility, symmetry etc.

This scheme operates on the idea of dividing or partitioning the large grid into smaller 2x2 cells. These smaller cells are now used as subjects to apply the automata's rule. The dotted and solid sub-lattices are used at odd and even time steps respectively, during evolution of the cellular automata.

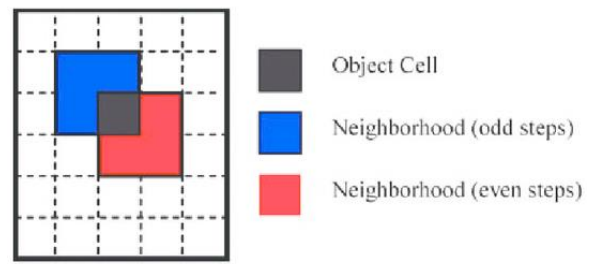

Fig. 1  Margolus Neighborhood Schematic.

A special transition function is used in the CA which reverses the state of every cell in a Margolus block consisting of $2x2$ cells, except for a block with exactly two live cells that remain unchanged. Blocks with 3 live cells experience a 180 degree rotation and the states of the 4 cells are flipped. This transition function has a distinctive property of being reversible. This makes the Critter's Rule a Reversible Cellular Automaton (RCA). Figure 2 depicts the Critter's rule which although has sixteen possible variations but only the representative of each case has been shown here, considering that this rule is rotation invariant.

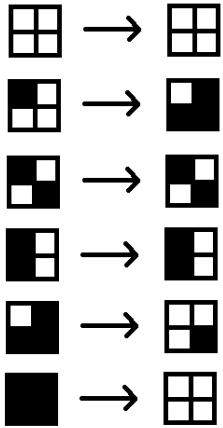

Fig. 2 Transformation function for Critter's rule

In modern cryptography, a trapdoor function is defined as a function that is easy to perform in one direction, yet difficult (impossible) to compute in the opposite direction (inverse) efficiently without knowing some special information. In mathematical terms, if f is a trapdoor function, then $y = f(x)$ is easy to calculate but $x = f^{-1}(y)$ is tremendously hard to compute without some special knowledge k (called key). In case k is known, it becomes easy to compute the inverse $x = f^{-1}(x, k)$. The entire field of cryptography relies on these special mathematical trapdoor functions that make it impossible for an eavesdropper to gain access to classified information while ensuring that the authorized parties (who know the secret key) trying to communicate can continue sharing information securely.

Reversibility is the key point of the construction of the trapdoor function. Partitioning neighborhoods as mentioned earlier is one of the common techniques employed to explicitly model a reversible CA.[11] But, in order to obtain a reversible cellular automaton, it is imperative that the rule for evolving each block must also be reversible. This can be explained by considering two separate configurations of a block, namely A and B which steer to the same result state C. Then a global configuration with A in a single block would be unidentifiable after one iteration of the configuration where the block state changes from A to B.

Thus, the proposed cryptosystem utilizes the reversibility of this cellular automaton as an innovative trapdoor function that can ensure a secure transmission of sensitive data and prevent any sort of penetration attacks.

## 4. The Proposed Cryptosystem

4.1 Encryption
a. The input string (plaintext) entered by the user is split into blocks based on the specified block size which can be adjusted for the desired level of security. The last block undergoes padding with a fixed predefined character.
b. Including the padding character, all characters in each block are converted to their corresponding ASCII values.
c. Each ASCII value is then converted to binary (base 2) representation. Since the largest value present on the ASCII table is $128$, all ASCII values are converted to $8$−bit long binary equivalents with the right number of 0s added in front.
d. At this stage, each block can be represented as a $2$−dimensional array consisting of 0s and 1s.
e. For each block (now represented as lattice), the Margolus Neighborhood is taken into consideration and evolved. It uses a partitioning scheme where the lattice is divided in isolated blocks of size $2x2$. Each $2x2$ block moves down and to the right with the next generation, and then moves back. Note that at each timestep during evolution, the transformation rule (Critter's rule) are applied on each of the $2x2$ neighborhoods.
f. The $2D$ lattice is evolved iteratively $K$ number of times where $K$ is the private key chosen by the user.
g. Each row of the evolved grid is a binary number which is now converted back to the decimal representation. These decimal values can be anything from $0$ to $255$.
h. The decimal values in the range $[0,255]$ are converted to the corresponding greyscale values ($0$ being black and $255$ white) and saved as an image. This is the encrypted image generated for the plaintext.
i. Each block is converted into such images and stacked together.

4.2 Decryption
a. The pixel values are read from the encrypted images and converted to their $8$−digit long binary equivalents. Each image is now a $2D$ lattice of $8x8$ dimension consisting of 0s and 1s.
b. The Margolus neighborhood is once again taken into consideration and this time, for each of the $2x2$ cells, the inverse transformation is applied for obtaining the new state of the neighborhood. Each $2x2$ block moves down and to the right with the next generation, and then moves back.

c. The above step is applied iteratively for $K$ times for each lattice where $K$ is the private key.
d. For all lattices, each row is converted back to decimal representation (i.e. ASCII values) from the 8-bit binary code.
e. All ASCII values obtained in the previous step are converted back to character notation and concatenated.
f. The padding characters are snipped off from the last block and the original plaintext is retrieved.

## 5. Proposed System Architecture

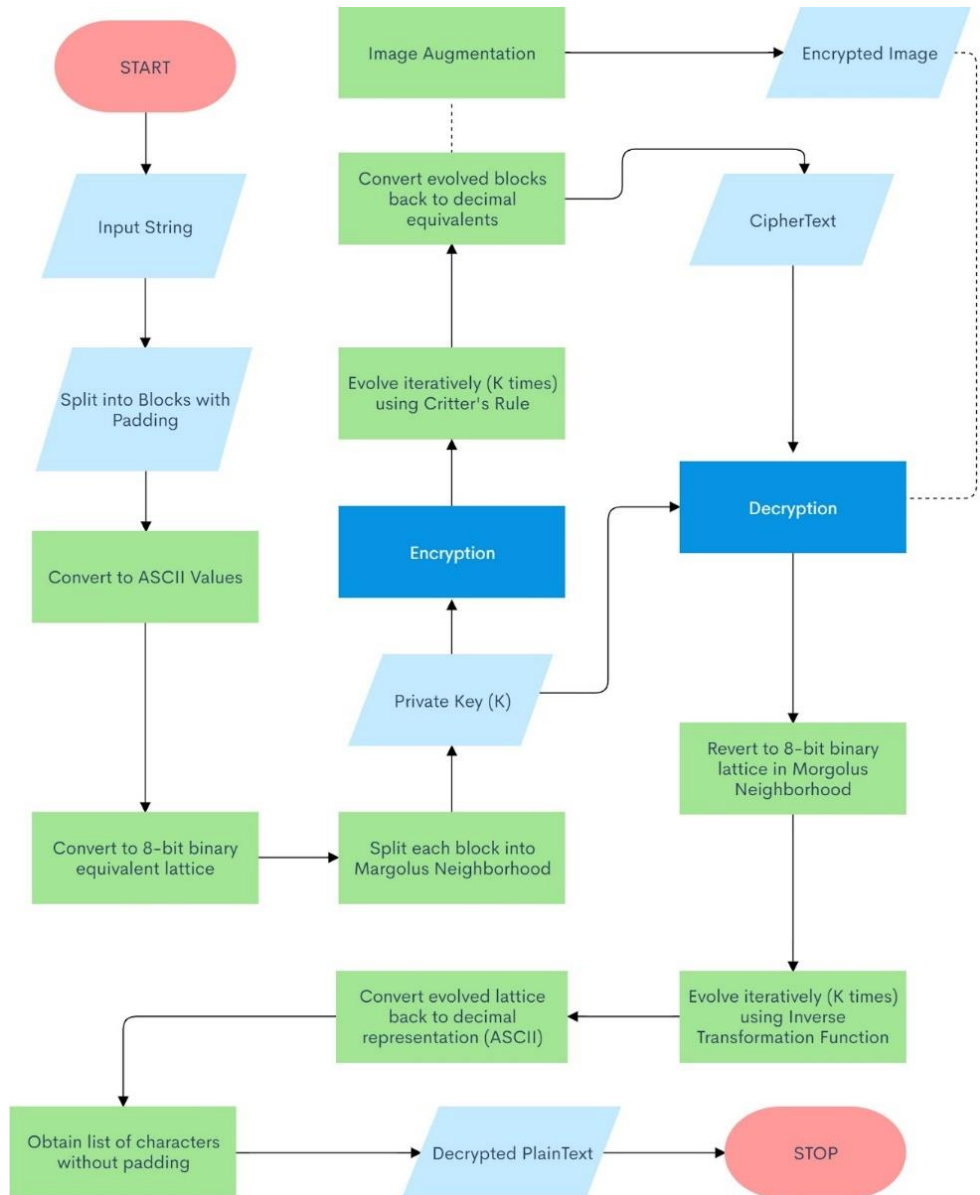

Fig. 3 Algorithm Block Schematic

## 6. Observed Security Features

The security of the proposed algorithm was put to test by computing the levels of confusion and diffusion in the generated ciphertext. Two simple messages ('$abcd$' and '$zbcd$') were considered. Note that both messages are identical except for one single character ('$a$' and '$x$'). The block size and private key were set to 8 and 4 respectively. Upon running the encryption algorithm involving the RCA (evolved for just 4 iterations), the following ciphertexts were obtained:

$‘abcd’ \to [[64, 49, 145, 19, 35, 70, 38, 33]]$
$‘abcx’ \to [[161, 21, 17, 39, 140, 32, 32, 179]]$

If a comparison is made between each element of both the ciphertexts, one can easily notice that there is 0% similarity. This implies that cryptanalysis techniques which rely on similarity will fail to crack this cryptosystem. This attribute proves the fact that even with a very few iterations, the cryptosystem is still able to deliver a feasible degree of security against attacks.

Another interesting attribute of this cryptosystem is that if the same character appears in a string multiple times, it gets encrypted to distinct numbers that do not follow any observable pattern or sequence. For instance, if the plaintext "$aaaaa$" is fed into the cryptosystem, it returns the following ciphertext for $blocksize = 8$ and $private\ key = 12$:

$"aaaaa" \to [[138, 11, 80, 87, 96, 44, 2, 18]]$

From an interceptor's perspective (even though the secret message is simply the character '$a$' written 5 times) the ciphertext will appear to be just a random sequence of numbers which makes it all the more difficult to come up with a logical approach to crack the secret message without any knowledge of the private key.

## 7. Image Augmentation

One method for sharing the encrypted data, other than simply sending the character list of mutated data, is sending it via an image file.

Here the data has been broken down into chunks and represented as $2D$ matrices. The range of values of this $2D$ matrix varies between 0 and 255. This is the exact range of values that is conventionally used to represent values of pixels across different channels.

Therefore, to augment the encrypted data in the form of an image, all we need to do is project the values of this $2D$ matrix to pixel values to obtain a row grayscale image. This row image is then extrapolated to obtain a square image that represents the data of one chunk.

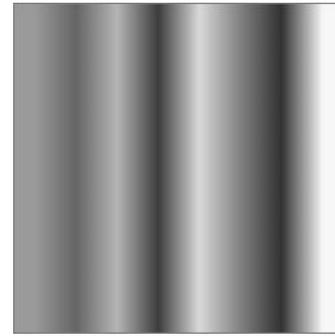

Fig. 4  Image augmented for a chunk of size 8, resized to 256 for clarity

For larger data, that exceeds the length of the data chunk that we consider in this algorithm, we will clearly obtain multiple $2D$ matrices and hence multiple images. The data from these images are therefore stored in a .h5 file. This data can be later read by the targeted receiver.

## 8. Performance Analysis

8.1 String Length vs time with fixed number of iterations (private keys)

For this analysis, the number of iterations for evolving the lattice was set to 8, 32, 64 and the length of the plaintext was gradually increased in steps of powers of 10 while logging down the time it takes to complete each successful encrypt-decrypt cycle. The plaintext was split into blocks of 8 characters and appropriate padding was applied before encrypting.

Table 1: String Length vs. Runtime with fixed Private Keys

| String Length | Encrypt-Decrypt Runtime (seconds) | | |
|---|---|---|---|
| | 8 iterations | 32 iterations | 64 iterations |
| 10 | 0.002 | 0.006 | 0.008 |
| 100 | 0.007 | 0.027 | 0.051 |
| 1000 | 0.078 | 0.235 | 0.461 |
| 10000 | 0.639 | 2.373 | 4.712 |
| 100000 | 6.475 | 23.731 | 48.477 |

The following graph shows that using smaller number of iterations ensures that the entire encrypt-decrypt operation takes place in a reasonable amount of time.

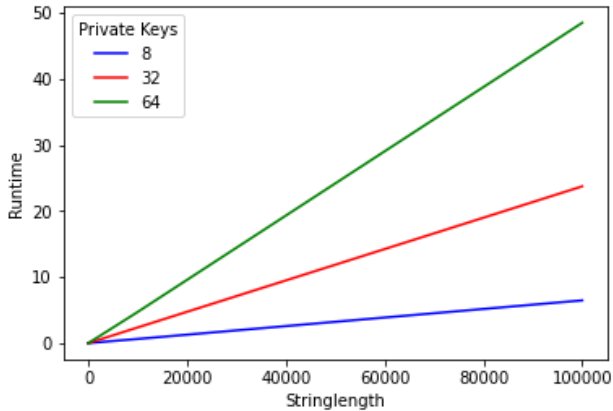

Fig. 5  String Length vs. Runtime for fixed number of RCA iterations .

## 8.2 Number of Iterations vs Runtime with fixed String

A test was conducted to study how varying the number of iterations of evolving the cellular automata has an impact on the time it takes for successful encryption followed by decryption of a predefined string. A fixed string of length 100000 characters (including alphabets and numeric values) was considered for this analysis and the number of iterations was gradually increased from 1 to 64. The following table shows the results obtained for a block size of 8 characters.

Table 2: Iterations (public key) vs. Runtime with String of length 1000000 characters

| Number of Iterations | Runtime (seconds) |
|---|---|
| 1 | 0.149 |
| 2 | 0.209 |
| 4 | 0.362 |
| 8 | 0.679 |
| 16 | 1.303 |
| 32 | 2.513 |
| 64 | 5.040 |

The proposed algorithm assumes that the $2D$ lattices for each block wraps around the edges. Due to this concept, it has been observed that using fewer iterations for evolving the lattices is more than sufficient to introduce a very high degree of security while ensuring that only a negligible amount of time is taken for the entire encryption-decryption operation.

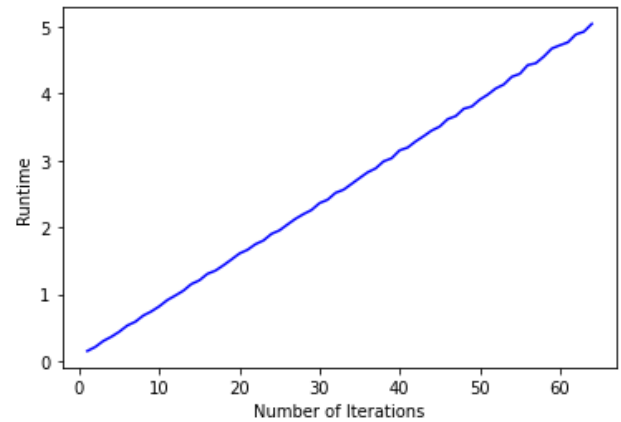

Fig. 6  Public Key (no. of iterations of RCA) vs. Runtime for fixed string of length 100000 characters.

## 9. Future Work

For the moment, the work done so far has been compiled into a GitHub Repository [12]. There is more scope for research on how well this cryptosystem will hold up against attacks from a quantum algorithm. Since this algorithm uses bit manipulation on a 2D lattice, it can very efficiently be implemented on an FPGA board to practically test out the multi-processing and multi-threading capabilities of this algorithm [7]. On machines having multi-core processors or GPUs, each $2x2$ neighborhood of the RCA can be simultaneously assigned and evolved by individual threads. Considering the different versions of Critter's rule, it might be possible to generate a more secure key by appending the rule as a code, with the number of iterations. Further research may be done by using other RCA that might be computationally lighter and/or provide a higher level of security - in other words, better suited automata for encryption than the one used in this paper.

## 10. Conclusions

The idea of this paper is to introduce an original cryptosystem that utilizes an unconventional trapdoor function which is built upon the concept of Reversible

Cellular Automata. The proposed algorithm requires the selection of the block size and the iterations for evolving the RCA as the private key. On a machine that works on sequential processing, it has been observed that choosing fewer iterations for evolving the RCA leads to a drastic reduction in the time it takes to encrypt, and decrypt a given message. However, security is not compromised owing to the edge wrap-around feature of the $2D$ lattice that is evolved.

Transmitting data in the form of ciphertext is secure enough using the proposed algorithm. However, to add a layer of protection, the data can additionally be encoded as a grayscale image. The nonchalant nature of the data not only leads an unsuspecting interceptor astray, but also adds an extra layer of security.

With the recent advancements in modern computing, specifically in the area of processing power, many existing encryption algorithms have started falling short of the expected standards. Thus, there has risen a need to develop new and numerous algorithms that ensure data security. Our approach went towards Cellular Automata owing to their simple construction and efficient numerical implementation with reduced computational complexity, while encryption, compared to conventional methods.